\documentclass[a4paper,11pt]{article}
\usepackage{pos}
\usepackage{graphicx}
\usepackage{booktabs}
\usepackage{orcidlink}
\usepackage{lineno}

\title{Enhanced track reconstruction at the HL-LHC}
\ShortTitle{Enhanced track reconstruction at the HL-LHC}

\author*[a]{Brunella D'Anzi \orcidlink{https://orcid.org/0000-0002-9361-3142} } 
\onbehalf{on behalf of the CMS and ATLAS Collaborations}
\affiliation[a]{University of California San Diego, \\ 9500 Gilman Dr., La Jolla, CA 92093, United States of America}
\emailAdd{brunella.danzi@cern.ch}

\abstract{The High-Luminosity Large Hadron Collider (HL-LHC) will substantially increase the \mbox{charged-particle} multiplicity and detector occupancy in the ATLAS and CMS experiments, posing new challenges for the reconstruction of charged-particle tracks.~Both experiments are upgrading their tracking detectors, with fully silicon-based systems providing a finer granularity, extended acceptance and updated geometry with respect to the current ones, with the goal to reduce the combinatorial complexity of pattern recognition tasks.~At the same time, the more demanding reconstruction environment at the HL-LHC requires algorithms capable of retaining high tracking efficiency while reducing the associated computational cost.~To this purpose, both the ATLAS and CMS Collaborations are developing tracking algorithms for offline and online reconstruction, exploiting parallelized, vectorized and heterogeneous architectures.~The ATLAS experiment is moving toward an ACTS reconstruction chain, based on the Combinatorial Kalman Filter algorithm.~In the CMS experiment, the new High-Level Trigger tracking baseline combines the Patatrack and the Line Segment Tracking algorithms for seed reconstruction, using the mkFit algorithm for track building.~These developments yield tracking efficiencies above 85\% across the full phase space at high pileup, extending the acceptance to forward and displaced tracks with transverse displacements of up to tens of centimetres, while significantly reducing the reconstruction time.~Machine-learning techniques are also being explored at several stages of the reconstruction, including track building.~The state-of-the-art performances of the tracking strategies developed by the ATLAS and CMS Collaborations for the HL-LHC are presented, together with prospects of further improvements.}

\FullConference{14th Edition of the Large Hadron Collider Physics (LHCP2026)\\
18-22 May 2026\\
Paris, France\\}

\begin{document}

\maketitle

\section{Introduction}
The High-Luminosity Large Hadron Collider (HL-LHC) will open new opportunities for discovery and precision measurements, but it will also push the reconstruction of charged-particle tracks into a much more demanding regime.~The tracking challenge is especially severe because the average number of simultaneous proton--proton interactions per bunch-crossing (pileup, PU) is expected to increase from about 60 during the Run~3 of the LHC to about 200 in the HL-LHC.~This corresponds to an increase from $\mathcal{O}(10^3)$ to $\mathcal{O}(10^4)$ charged-particles within detector acceptance, with $\mathcal{O}(10^5)$ hits (or space-points) per event.~The resulting combinatorics directly increases the computational complexity of the track reconstruction task.~Pileup also has a physics impact.~Pileup and misreconstructed tracks can degrade the hard-scatter vertex and jet-energy reconstruction, thereby reducing the sensitivity of physics analyses. 

The CMS and ATLAS experiments address this problem through a combination of detector and algorithm upgrades.~New silicon trackers provide finer granularity and redundancy.~The new ATLAS tracker will consist of an Inner Tracker, with pixel coverage up to $|\eta| = 4$ \cite{ATLAS_ITk_Pix} and a strip detector with double-sided microstrip modules up to $|\eta|=2.7$ \cite{ATLAS_ITk_St}.~The new CMS tracker will consist of a densely segmented pixel Inner Tracker (IT) with coverage up to $|\eta| = 4$, and an Outer Tracker (OT) with coverage up to $|\eta| = 2.5$.~In the OT, closely spaced macro-pixel and/or strip sensors can directly provide an estimate of the transverse momentum ($p_{\mathrm{T}}$) of the tracks \cite{cms-tracker}.~Dedicated precision-timing systems will add a fourth-dimensional handle against pileup to improve track-to-vertex association:~the ATLAS High Granularity Timing Detector will cover $2.4<|\eta|<4$ \cite{ATLAS_HGTD},~while the CMS Minimum Ionizing Particle Timing Detector (MTD) will cover up to $|\eta|=3$ \cite{CMS_MTD}.

The reconstruction chain follows a similar strategy in the two experiments: exploit the upgraded detector information to build high-quality tracks, and offload computationally expensive operations to architectures that can exploit modern CPUs and GPUs. 

\section{Track reconstruction at ATLAS and CMS at the HL-LHC}
The track reconstruction at ATLAS and CMS at the HL-LHC is expected to be an iterative procedure \cite{cms-tracking,atlas-itk}.~High-quality prompt tracks are reconstructed first;~their hits are removed, and subsequent iterations use the remaining hit measurements to recover lower-momentum, detached, or otherwise difficult tracks.~Dedicated steps address special regions of phase space, such as dense jet environments or photon conversions.~Using an iterative procedure reduces the combinatorics for later iterations, with a consequent gain in performance.

The track reconstruction at the High-Level Trigger (HLT) \cite{cms-hlt,atlas-ef-tracking}, must satisfy more stringent computational constraints.~In this report, both ATLAS and CMS consider the starting point of the HLT track reconstruction to be equivalent to a truncated version of the offline reconstruction sequence, with only the first two iterations performed.

After the local hit reconstruction, the tracking chain in each iteration can be separated into four stages: seeding, track building, fitting, and selection.~In the seeding step, compatible hits are grouped into short objects that are used as the starting point for pattern recognition, or track building.~At the HLT, the ATLAS experiment is planning to move towards the usage of pixel triplets for the main iteration, and strip space-points for the second Large-Radius Tracking (LRT) iteration~\cite{ATLAS_ACTS_LRT}.~In CMS, the current plan for HLT is to use pixel quadruplets including IT and OT hits produced with the Patatrack algorithm  \cite{patatrack} together with OT-only seeds produced with the Line Segment Tracking (LST) algorithm \cite{lst,lst-2,lst-3}.~Both Patatrack and LST are heterogeneous algorithms, implemented using the Alpaka hardware-portability framework \cite{alpaka,alpaka-tuning,Worpitz2015}, enabling their execution on both CPUs and GPUs.

Track building extends the seeds by adding compatible hits on subsequent detector layers.~The ATLAS Collaboration uses a single Combinatorial Kalman Filter (CKF) \cite{Fruehwirth_Kalman} iteration on CPU within the ACTS framework \cite{Ai_ACTS}.~The CMS experiment uses the mkFit algorithm \cite{parallel-kalman,mkfit}, a vectorized and parallelized implementation of the CKF algorithm designed to reduce the cost of track propagation on CPUs.

The fitting stage extracts the track parameters and rejects incompatible measurements.~In the presented ATLAS configuration, the parameters are obtained directly from the CKF building step, without a separate fit.~In CMS, a CKF fit is used, with the plan to exploit the mkFit fitting implementation~\cite{CMS-DP-2026-031} in the upcoming future.~Track selection then ranks candidates using the hit multiplicity and the track quality information.

In the following, the proposed solutions are compared to the “legacy” workflows. In ATLAS, the legacy algorithm consists of a non-ACTS implementation using both pixel and strip seeds, followed by a global $\chi^2$ fit. In CMS, two CKF-based iterations are used, seeded with pixel IT-only quadruplets and triplets, respectively.

\begin{figure}[htp!]
	\centering
	\includegraphics[width=0.45\textwidth]{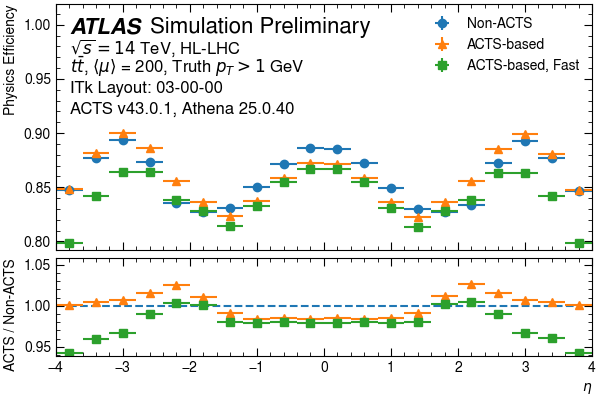}
	\includegraphics[width=0.4\textwidth]{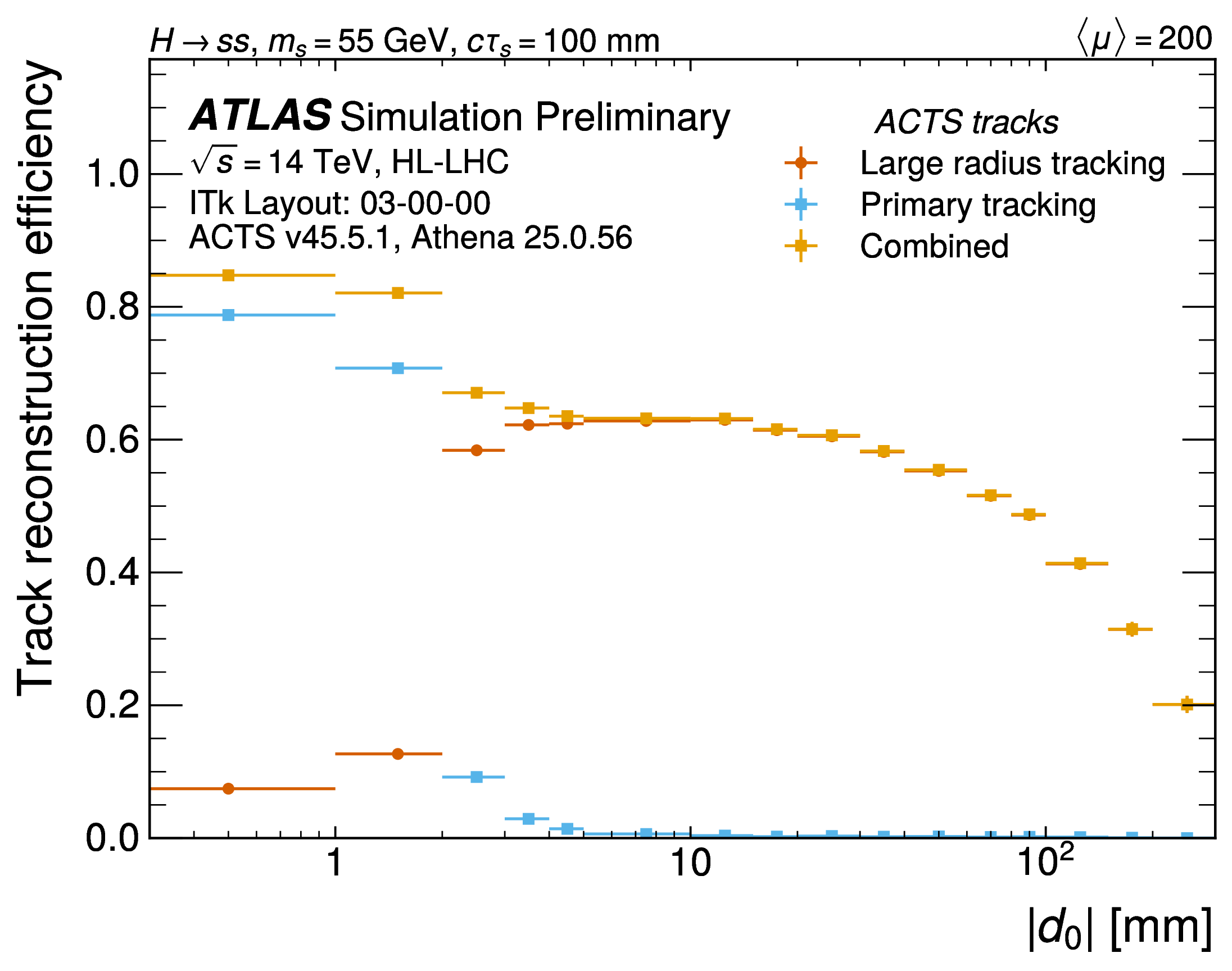}
	\caption{The HL-LHC track reconstruction efficiency in ATLAS: (left) as a function of the simulated track pseudorapidity $\eta$, for legacy (non-ACTS:~blue) and ACTS (offline-like:~orange;~HLT-like:~green) configurations from Ref.~\cite{ATLAS_ACTS_ITk}; (right) as a function of the absolute value of the simulated track transverse impact parameter, for the primary tracking iteration (blue), the Large-Radius tracking iteration (red), and their combination in the offline reconstruction (orange) from Ref.~\cite{ATLAS_ACTS_LRT}.}
	\label{fig:fullTrackingATLAS}
\end{figure}

\begin{figure}[htp!]
	\centering
	\includegraphics[width=0.45\textwidth]{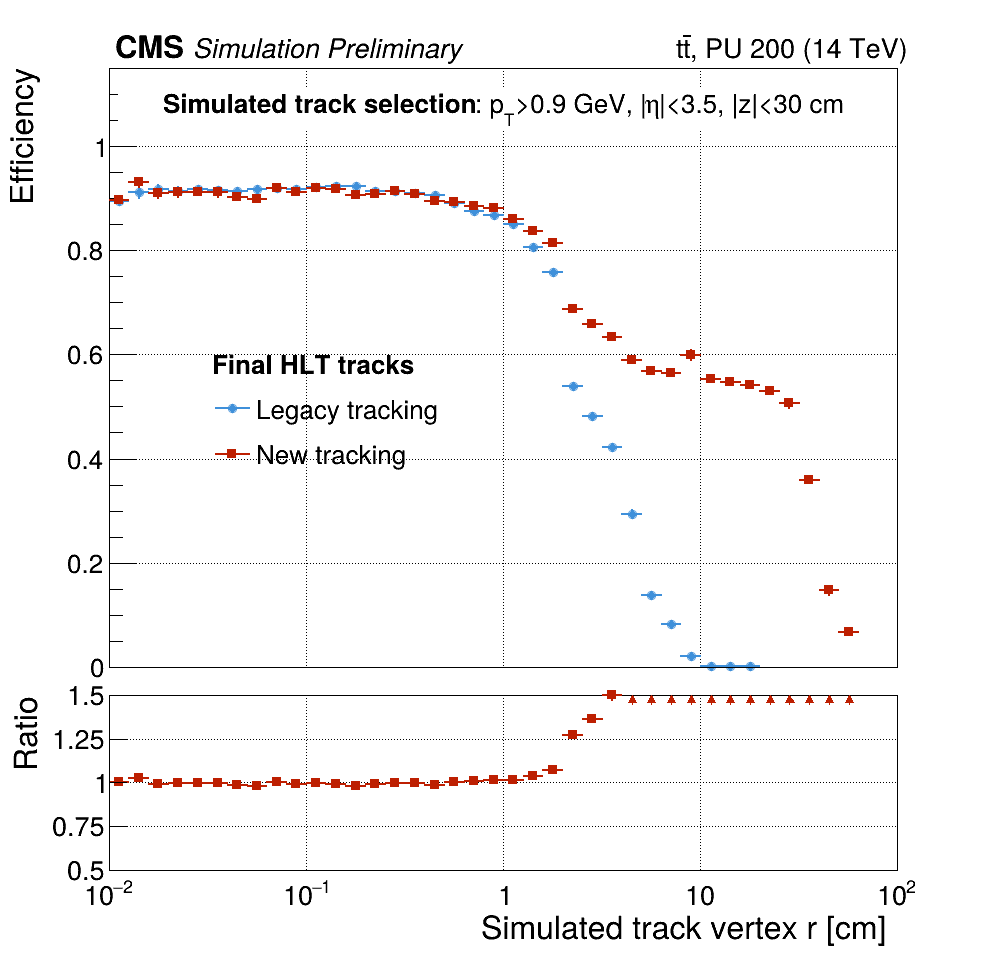}
	\includegraphics[width=0.45\textwidth]{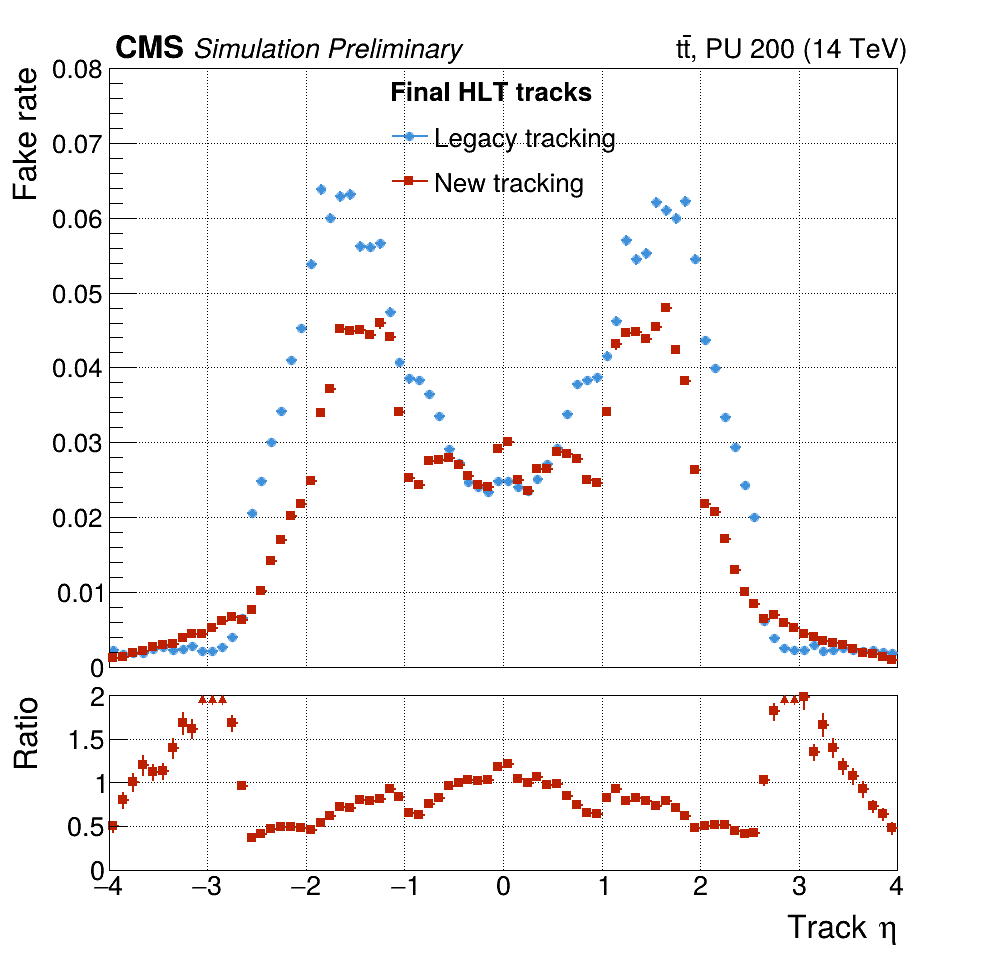}
	\caption{(Left) The HL-LHC track reconstruction efficiency at CMS HLT,~as a function of the simulated track production vertex radial displacement.~(Right) Fake rate for reconstructed tracks,~as a function of the reconstructed track pseudorapidity $\eta$, for the legacy (blue) and newly proposed (red) configurations~\cite{lst-3}.}
	\label{fig:fullTrackingCMS}
\end{figure}

\section{Tracking physics and timing performance at the HL-LHC}

The tracking physics performance is evaluated in terms of tracking efficiency and fake rate;~efficiency is defined as the fraction of simulated tracks from the hard interaction vertex that are associated to a reconstructed one, while fake rate is defined as the fraction of reconstructed tracks that are not associated to any simulated track.~The association of simulated and reconstructed tracks is based on the fraction of shared hits, with the specific definition differing between the two experiments, as described in Refs. \cite{ATLAS_ACTS_ITk,cms-tracking}.

The tracking efficiency for the newly proposed ATLAS configuration for HL-LHC, compared to the legacy configuration, is shown in \autoref{fig:fullTrackingATLAS} \cite{ATLAS_ACTS_ITk,ATLAS_ACTS_LRT}.~On the one hand, the main ACTS-based iteration is shown to be consistent within a few percent with the legacy tracking,~both for the offline and HLT reconstruction.~On the other hand, the second LRT offline tracking iteration efficiently reconstructs tracks with transverse impact parameters up to 30 cm. The overall fraction of fake tracks is $\mathcal{O}(10^{-4})$.~For the reconstructed tracks, the smaller pixel pitch and better resolution in the bending plane provided by the silicon microstrips contribute to the improvement of impact parameter and momentum resolutions, as well as of the tracking performance in dense environments~\cite{ATL-PHYS-PUB-2026-002}.

The tracking physics performance for the CMS newly proposed configuration at the HLT \cite{lst-3} is shown in \autoref{fig:fullTrackingCMS}.~The achieved tracking efficiency is similar to that of the legacy tracking, except for large displacements, where the new configuration efficiently extends the acceptance up to production vertex radial displacements of 60 cm.~At the same time, the overall fake rate is reduced already at the level of seed tracks.~Track resolutions are comparable to those obtained with the legacy tracking \cite{lst-3}.

In terms of timing, the newly proposed ATLAS configuration shows a significant timing improvement, with a 50\% reduction in CPU time for the main HLT reconstruction step \cite{IDTR-2025-06};~offloading to GPU in ATLAS is currently under study \cite{ATL-DAQ-PUB-2025-00}. The CMS newly proposed configuration also improves the track reconstruction time by about 35\% compared to the legacy algorithm when executed only on CPUs, and by 64\% with GPU offloading.~This corresponds to reductions of the total event reconstruction time by 9\% and 33\%, respectively \cite{lst-3}.~An additional reduction of about 70\% of the fitting step could be achieved when using the mkFit algorithm for track fitting, too  \cite{CMS-DP-2026-031}.
\section{Machine-learning techniques for tracking and vertexing}
Machine-learning techniques are being integrated or investigated at several levels in both experiments.~For example, CMS employs shallow neural networks in its GPU-based reconstruction to reduce the fake rate, and learned track embeddings to remove duplicate LST objects and improve the reconstruction efficiency for highly-displaced tracks \cite{machinelearningcms-2}.

\begin{figure}[htp!]
	\centering
	\includegraphics[width=0.45\textwidth]{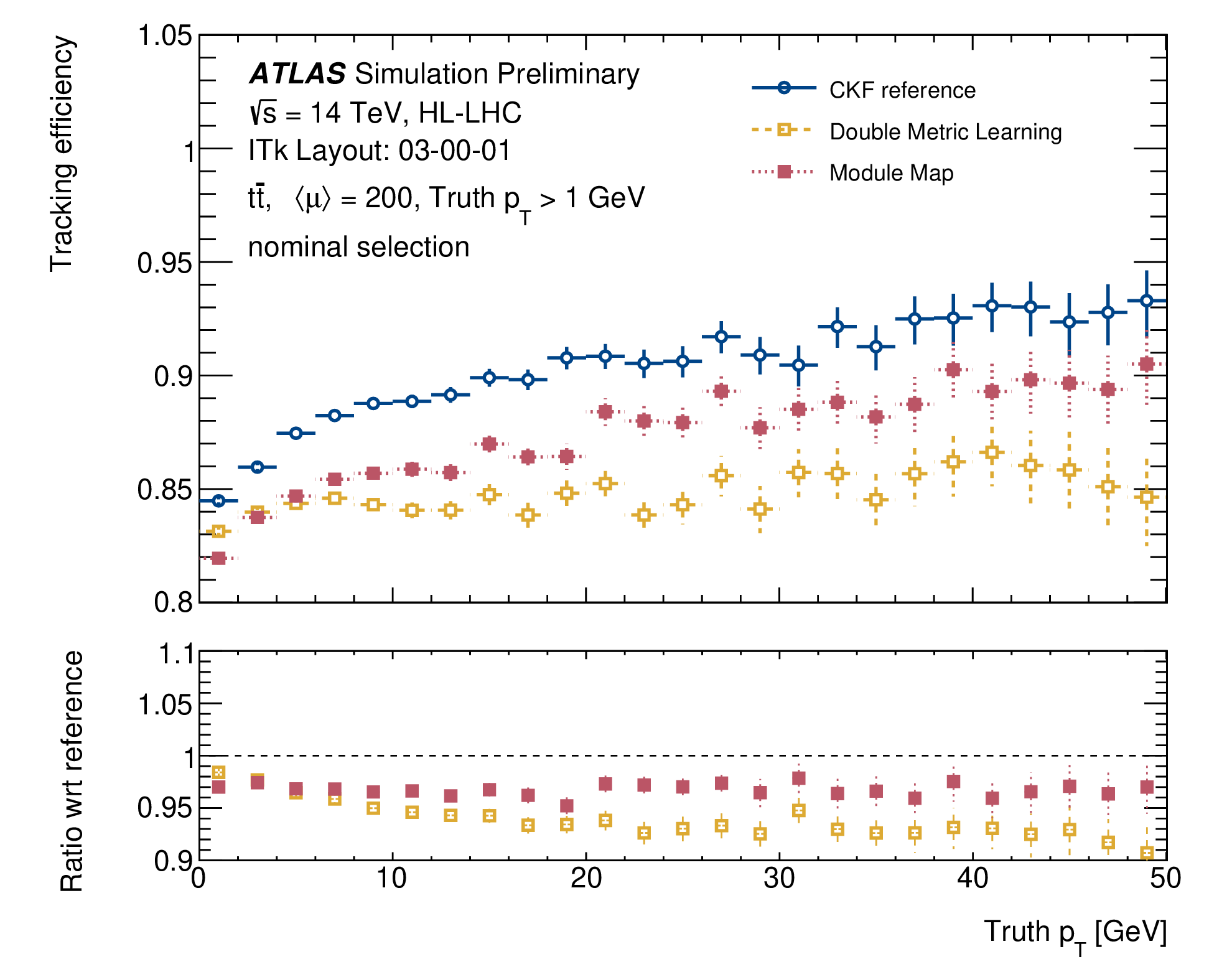}
	\includegraphics[width=0.5\textwidth]{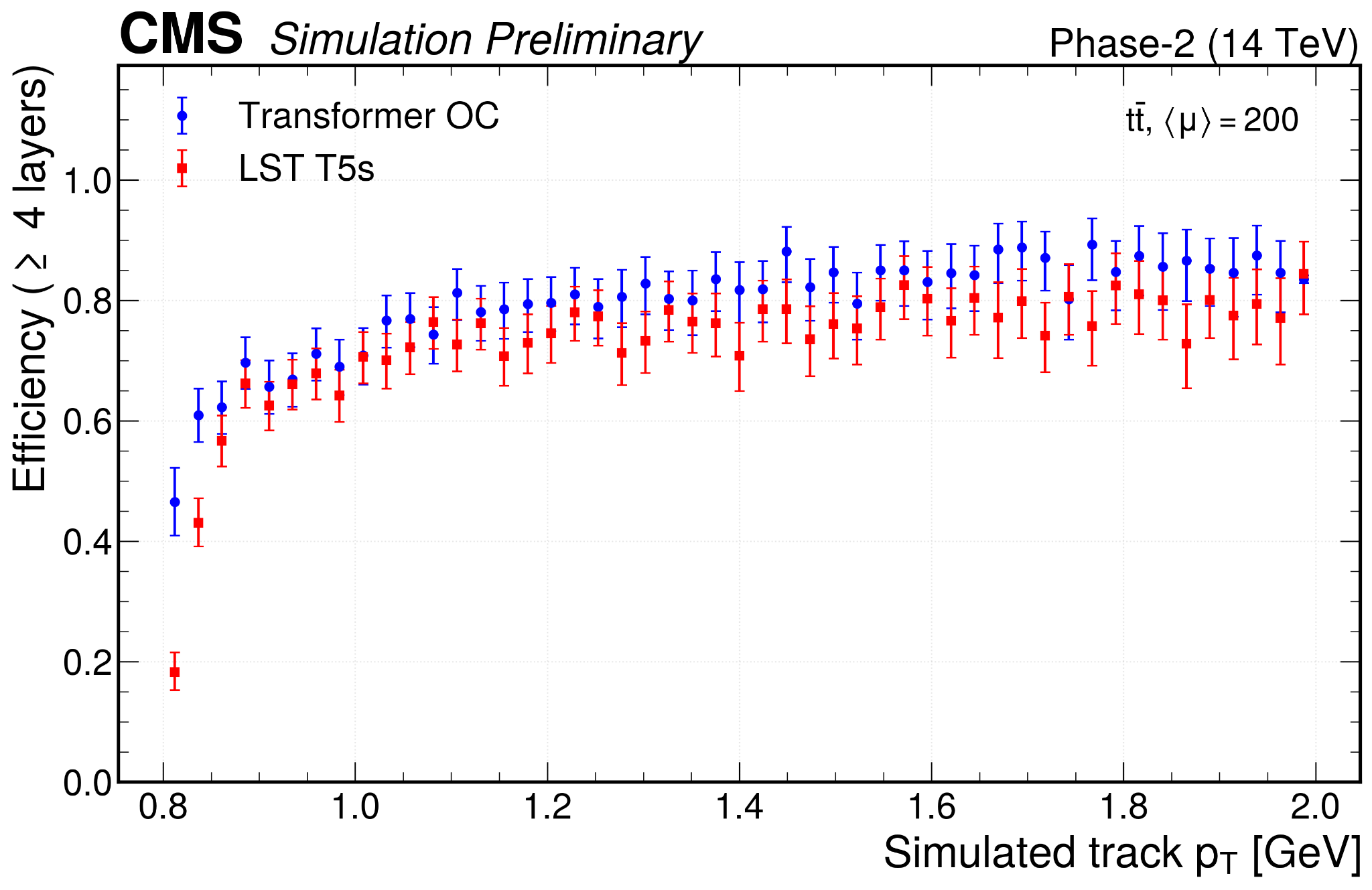}
	\caption{(Left) The HL-LHC tracking efficiency in ATLAS, as a function of the simulated track $p_{\mathrm{T}}$, achieved with the legacy reconstruction algorithm (blue) and with different configurations of the GNN4ITk algorithm (yellow and red) \cite{machinelearningatlas}.~(Right) The HL-LHC tracking efficiency in ATLAS, as a function of the simulated track $p_{\mathrm{T}}$, achieved with the LST T5 objects (red) and with the transformer model using LST T3 objects as inputs (blue) \cite{machinelearningcms}.}
	\label{fig:MachineLearning}
\end{figure}

Both the ATLAS and CMS Collaborations are also investigating graph neural networks and particle-transformer architectures for track building from space-points \cite{machinelearningatlas} and the LST objects \cite{machinelearningcms}, respectively.~As shown in \autoref{fig:MachineLearning}, these approaches achieve physics performances close to those of the corresponding benchmark configurations.

Work is ongoing also on the development of vertex reconstruction algorithms, in both experiments.~In CMS,~recent progress was shown on both vertexing algorithms that only rely on tracker-only information \cite{patatrack}, and algorithms that also exploit the timing information from the MTD, either relying on the traditional deterministic annealing method \cite{vertexMTD}, or on graph neural networks \cite{CMS-DP-2026-042}.~In ATLAS, efforts are ongoing to improve the Run-3 MultiVertex Fitter strategy \cite{ATL-PHYS-PUB-2019-015}, while exploring deep-learning-based alternatives \cite{ATL-PHYS-PUB-2023-01,ATLAS-IDTR-2025-03}.

\section{Conclusions and outlook}
The HL-LHC tracking strategies for the ATLAS and CMS experiments rely on the optimized use of detector information, reconstruction algorithms and computing resources.~The upgraded ATLAS and CMS trackers, together with redesigned reconstruction algorithms and heterogeneous computing models, provide high efficiency at large pileup, reduced fake rates, improved sensitivity to displaced tracks, and substantial computation timing gains.~The next steps include the finalization of the ATLAS migration to ACTS, continued optimization of the CMS track reconstruction tasks, and further studies of detector-failure scenarios.~Machine-learning approaches show promising results for track building, however further development and validation are required before considering a deployment in production.

\section*{Acknowledgements}

The author acknowledges the support by the National Science Foundation under Cooperative Agreements OAC-1836650 and PHY-2323298 within the CMS Collaboration work.

\end{document}